# Quantifying the relevance of different mediators in the human immune cell network

P. Tieri[1], S. Valensin[2], V. Latora[3], G.C. Castellani[2], M. Marchiori[4], D. Remondini[2], C. Franceschi[1,2,5].

[1] *Dipartimento di Patologia Sperimentale, Università di Bologna, Via San Giacomo 12, 40126 Bologna, Italy.*

[2] *C.I.G. – Centro Interdipartimentale "L. Galvani" per Studi Integrati di Bioinformatica, Biofisica e Biocomplessità, Università di Bologna, Via Tolara di Sopra 50, 40064 Ozzano Emilia, Italy.*

[3] *Dipartimento di Fisica ed Astronomia and INFN - Istituto Nazionale Fisica Nucleare, Università di Catania, Via S. Sofia 64, 95123 Catania, Italy.*

[4] *W3C MIT Lab for Computer Science, 545 Technology Square, Cambridge, MA 02139, USA and Dipartimento di Informatica, Università di Venezia, Via Torino 155, 30172 Mestre, Italy.*

[5] *INRCA - Istituto Nazionale di Ricovero e Cura Anziani, Dipartimento Ricerche, Via Birarelli 8, 60100 Ancona, Italy.*

*Abstract*

Immune cells coordinate their efforts for the correct and efficient functioning of the immune system (IS). Each cell type plays a distinct role and communicates with other cell types through mediators such as cytokines, chemokines and hormones, among others, that are crucial for the functioning of the IS and its fine tuning. Nevertheless, a quantitative analysis of the topological properties of an immunological network involving this complex interchange of mediators among immune cells is still lacking. Here we present a method for quantifying the relevance of different mediators in the immune network, which exploits a definition of centrality based on the concept of efficient communication. The analysis, applied to the human immune system, indicates that its mediators significantly differ in their network relevance. We found that cytokines involved in innate immunity and inflammation and some hormones rank highest in the network, revealing that the most prominent mediators of the IS are molecules involved in these ancestral types of defence mechanisms highly integrated with the adaptive immune response, and at the interplay among the nervous, the endocrine and the immune systems.

---

Network analysis has emerged as a powerful approach to understanding complex phenomena and organization in social, technological and biological systems[1-4]. In particular it is increasingly recognized the role played by the topology of cellular networks, the intricate web of interactions among genes, proteins and other molecules regulating cell activity, in unveiling the function and the evolution of living organisms[5-9]. The cells of the IS have various ways to communicate among each other: directly, by establishing bounds between cell surface ligands and receptors, and indirectly, by means of a variety of soluble mediators released and bound by the immune cells[10-11]. Soluble mediators implement cellular communication both at short range (autocrine and paracrine cell stimulation) and across the major body systems (immune, endocrine and nervous systems). These mediators have a fundamental role in regulating the IS reaction to a possible danger, which triggers an integrated response involving both innate and clonotypic immunity, and which eventually results in an inflammatory response. Mediators are characterized by pleiotropy (each mediator has multiple targets) and redundancy (each mediator is produced by several sources), two characteristics that strongly influence the reliability of the IS, and that significantly contribute to its robustness and adaptability. From an experimental as well as theoretical point of view the main attention has been generally focussed on few limited subsets of immune cell types and mediators. Here we follow a different approach trying to grasp the global properties of the IS network, by modelling the whole system of immune cells as networked by soluble mediators. The network we consider is constituted by various immune cell types, which can act as both sources and targets of the exchanged mediators. The immune cell network is represented as a valued directed[3] graph $G$, where the $N$ cell types considered are the vertices of the graph and the $M$ soluble mediators form its $K$ arcs: a directed arc from vertex $i$ to vertex $j$ is defined by the existence of at least one mediator secreted by cell $i$ and affecting cell $j$. Cell self-stimulation by soluble mediators (autocriny), which is an important peculiarity of the immune cell network, is also taken into account. The value $e_{ij}$ attached to the arc is assumed to be equal to the number of different mediators connecting the cell $i$ to the cell $j$ (Fig. 1). We consider such a number as a measure of the importance of the communication between two cells along the arc, hence modelling this as an efficiency[12] that measures the bandwidth.

The IS is basically a parallel working system: all cells concurrently send information along the network through their outgoing arcs, i.e. by secreting mediators, and receive information through their incoming arcs, i.e. by binding the mediators. Two cells/vertices in *G* can communicate through various paths, connecting them with different levels of efficiency: the efficiency of a path is the harmonic composition[13] of the efficiencies of the component arcs. We assume that the communication between vertices *i* and *j* takes the most efficient path, and the efficiency of such path is indicated by $\varepsilon_{ij}$ [14]. This approach represents an extension to valued networks of the shortest path assumption commonly used in social[3], biological[1,5-7] and communication/transportation[1,2,12] networks. We characterize the system global properties by defining the *network efficiency* as[12]:

$$E(G) = \frac{1}{N^2} \sum_{i,j \in G} \varepsilon_{ij}$$

Network analysts have used centrality as a basic tool for identifying key individuals in a network[3]. A variety of measures of centrality have been proposed over the years to quantify the topological importance of a node in a graph[3,15]. Differently, here we propose a method for quantifying the centrality of the various soluble mediators in the IS. The method is based on the concept of efficient communication over the immune cell network. The centrality of each mediator $\alpha$ ($\alpha = 1, ..., M$) is measured by its *network relevance* $r_\alpha$, defined as the relative drop in the network efficiency *E* caused by the removal of the mediator, namely:

$$r_\alpha = \frac{E(G) - E(G'_\alpha)}{E(G)} \qquad \alpha = 1,...,M$$

where $G'_\alpha$ is the graph obtained by removing mediator $\alpha$ from *G*. In fact, in our framework, the removal of a mediator weakens some of the values $e_{ij}$ attached to the arcs and, consequently, affects the communication between various couples of cells, decreasing some of the $\varepsilon_{ij}$ and thus the IS network efficiency. The method has been applied to human data retrieved by the Cytokine Reference Database[16], taking into account $N=19$ immune cell types (list in Fig.1) and $M=93$ soluble mediators (list in Fig. 3), selected among all available mediators as those connecting at least two different cell types from our immune cell list. The resulting graph $G$ is dense, having $N=19$ vertices and $K=316$ arcs (including self-connections) out of the 361 arcs of the full graph, and thus a pure topological analysis would be poorly significant. Therefore, we performed a more refined analysis by taking into account the strength of the interactions among the IS cells. The integer values $e_{ij}$ attached to the arcs range from 1 to 36, since there are up to 36 different mediators connecting a couple of cells. In Fig. 2 we plot the cumulative relevance distribution defined as $P(r) = \frac{M(r)}{M}$, where $M(r)$ is the number of mediators with relevance larger than $r$. For large values of the relevance, $P(r)$ shows a power-law behaviour $P(r) \approx r^{-(\gamma-1)}$ with a scaling exponent $\gamma = 2.8 \pm 0.1$.

The fat-tail in the mediators relevance distribution indicates that the universal scaling principles discovered in other biological networks[1,2,5,7] seem to be also fundamental ingredients of the human IS architecture. This shows that the concept of importance is selective and radical: immune cells form a highly inhomogeneous network in which a few soluble mediators play a central role in mediating the interactions between the different cell types.

The network relevance of each mediator is reported in Fig. 3.

The plot shows that only 3 mediators, TGF-β, MIP-1-α and -β (grouped as one mediator[16]) and TNF-α have network relevance larger than 0.5; 11 mediators have network relevance in the range [0.2, 0.5]; and, the remaining 79 have network relevance in the range [0, 0.2]. The sum of the network relevance of the first three mediators accounts for the 20.4% of total mediators relevance, while those of the second and third groups account for 34.7% and 44.9%, respectively. The unequal role played by the mediators in the efficient coordination of the immune network can be quantified by means of the Gini coefficient. The Gini coefficient *g* is a measure used in economics and ecology to describe size-wealth inequality in a population. It compares the Lorenz curve of a ranked empirical distribution, i.e. a curve that shows, for the bottom x% of individuals, the percentage y% of the total size which they have, with the line of perfect equality[17]. The coefficient g ranges from a minimum value of zero, when all individuals are equal, to a theoretical maximum value of 1 in a population in which every individual except one has a size of zero. The value we have found *g*=0.64 confirms the large skew of relevance.

The three most important mediators are pro- and anti-inflammatory molecules, and are involved in the communication among a large number of cell types, respectively 216, 224 and 120. The second group of mediators includes 9 pro- and anti-inflammatory cytokines/chemokines, one non-inflammatory cytokine (IL-7), and one neuro-endocrine hormone (VIP/PACAP). So, notwithstanding the fact that among all mediators taken into account those involved in the inflammatory process are only the 24%, the 86% (twelve out of the fourteen) top molecules are all inflammatory and they account for the 50% of all inflammatory mediators.

The analysis shows that mediators involved in innate immunity and inflammation have the most central role in the immune network. In the last decade, innate immunity and inflammation emerged as a central component of the capability of the immune system to sense "danger" signals and to start immune responses, in turn involving cells of adaptive immunity[18,19]. The importance of innate immunity and inflammation is reinforced by recent data indicating that survival at extreme ages, and conversely mortality caused by major age-associated diseases, are related to low or high level of inflammation ("inflamm-aging")[20-22], respectively. From this point of view it is challenging that several inflammatory mediators, that have been experimentally studied and found to have a major role in aging and/or longevity, such as TGF-$\beta$[23], TNF-$\alpha$[24],

IL-10[25], IFN-γ[26], IL-6[27], MCP-1[28], are included in the first two groups of relevant mediators, while two inflammatory cytokines, such as IL-1α and IL-1β, which do not seem to play a relevant role in these phenomena[29], appear in the group of lower rank mediators. Thus, mediators involved in innate immunity - the most ancestral branch of the immune system - and in highly conserved defence pathways such as inflammation, appears to give a substantial contribution to the efficient communication of the IS network. Finally, the presence of a neuro-endocrine hormone, ranking high in the list of immune network mediators, suggests that further investigations with a set of cells and mediators including those from the neural and the endocrine systems, can shed light on the likely close interplay among these three systems[30].

**Acknowledgements** This research has been partially funded by Cofin grant from Italian Ministry of University and Research (MIUR), EU grants FUNCTIONGE and T-CIA, and INFN.

**Correspondence** C.F. (clafra@alma.unibo.it), V.L (vito.latora@ct.infn.it), P.T. (p.tieri@alma.unibo.it).

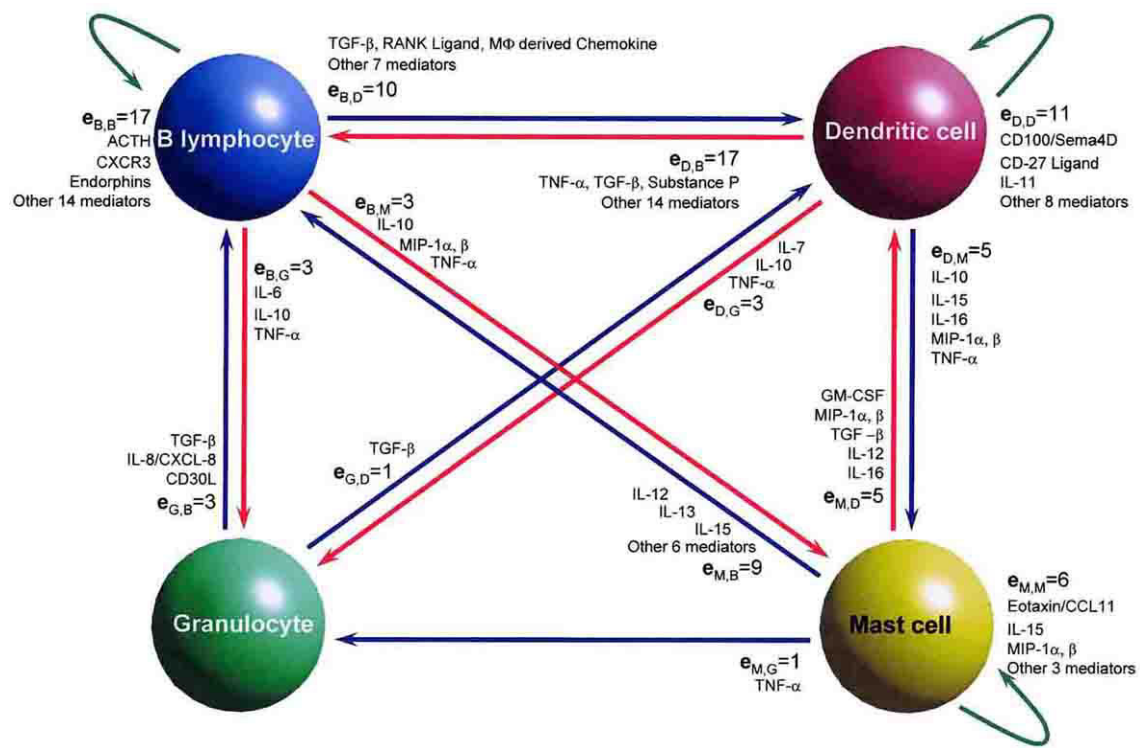

Figure 1. Portion of the immune cell network considered. In the figure a network with only four cell types out of the 19 considered, is illustrated. Cell types are the vertices of the graph, while diffusible mediators, which enable the cells to communicate each other, define the arcs of the graph. Autocriny is also considered and depicted in the figure. The arcs are shown in three different colours for image benefit. The value $e_{i,j}$ associated to each arc is equal to the number of different mediators connecting cell $i$ to the cell $j$. The name of the mediators are listed close to the respective arcs. The 19 cell types considered in the network are: B lymphocytes, basophils, dendritic cells, endothelial cells, eosinophils, epithelial cells, fibroblasts, granulocytes, macrophages, mast cells, monocytes, neutrophils, NK cells, spleen cells, stromal bone marrow cells, T lymphocytes, T CD4+ lymphocytes, T CD8+ lymphocytes, thymocytes.

(Legend for Figure 1. B: B lymphocytes; D: Dendritic cell; G: Granulocyte; M: Mast cell; MΦ: Macrophage)

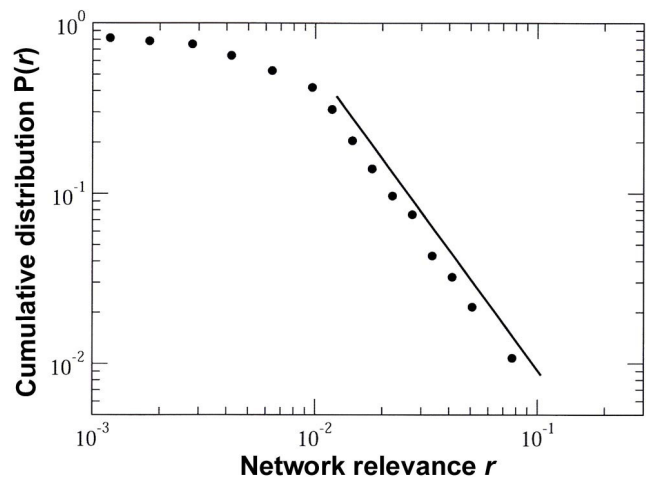

Figure 2. Distribution of the number of mediators having a network relevance larger than $r$. The relevance of a mediator is calculated as the relative drop in network efficiency caused by its removal from the network. To reduce noise, logarithm binning was applied[5]. The linear tail in the figure indicates a scale-free power-law behaviour in the relevance of the various mediators of the immune cell network that can be fitted with a curve $P(r) \cong r^{-(\gamma-1)}$ with $\gamma = 2.8 \pm 0.1$.

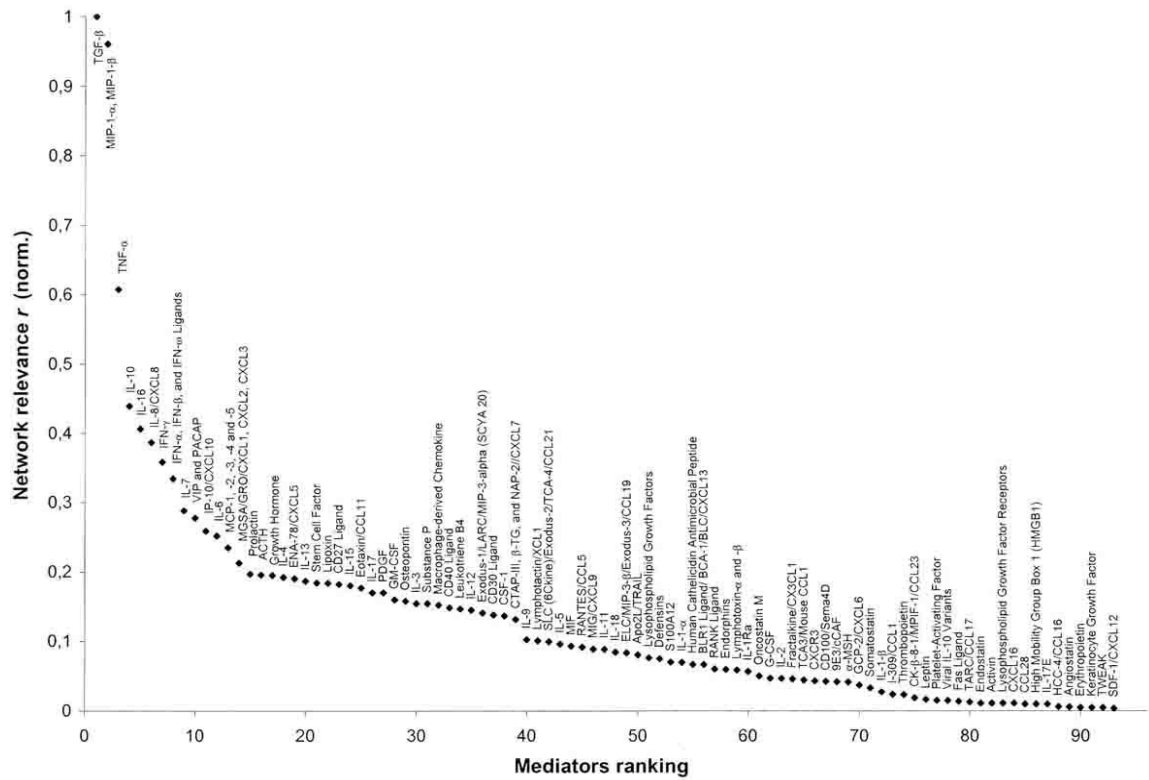

Figure 3. Network relevance *r* of the mediators of the immune cell network. Mediators are ranked on the basis of their *r* value, where values are normalized so that the maximum value is set to 1. The fourteen mediators in the highest ranks, i.e. those that play a crucial role for the network efficient communication, are all pro- or anti-inflammatory mediators, except IL-7 (a T cell growth factor) and VIP/PACAP (an immuno-neuro-endocrine mediator).